\documentclass[useAMS]{mn2e}
\usepackage{graphicx}
\usepackage{amsfonts}
\usepackage{txfonts}

\title[Galactic distribution of merging neutron stars and black holes]
  {Galactic distribution of merging neutron stars and black holes\\
   -- prospects for short $\gamma$-ray burst progenitors~and~LIGO/VIRGO}
\author[R. Voss and T.M. Tauris] 
  {R.~Voss\thanks{E-mail: voss@astro.ku.dk (RV); tauris@astro.ku.dk (TMT)} and T.~M.~Tauris$^\star$\\
  Astronomical Observatory, Niels Bohr Institute, University of Copenhagen, DK-2100 Copenhagen {\O}, Denmark}
\date{Received 2002 December 5; Accepted 2003 March 11}

\pagerange{\pageref{firstpage}--\pageref{lastpage}} \pubyear{2003}

\def\LaTeX{L\kern-.36em\raise.3ex\hbox{a}\kern-.15em
    T\kern-.1667em\lower.7ex\hbox{E}\kern-.125emX}

\begin{document}

\label{firstpage}

\maketitle

\begin{abstract}
   We have performed detailed population synthesis on a large number ($2\times10^7$) of binary systems
in order to investigate the properties of massive double degenerate binaries.
We have included new important results in our input physics 
in order to obtain more reliable estimates of the merging timescales and relative formation rates. 
These improvements include refined treatment of the binding energy in a common envelope, helium star evolution
and reduced kicks imparted to newborn black holes.
   The discovery and observations of GRB afterglows and the identification of host galaxies have allowed
comparisons of theoretical distributions of merger sites with the observed distribution of afterglow positions
relative to host galaxies.
To help investigate the physical nature of short- and long-duration $\gamma$-ray bursts (GRBs), we compute
the distances of merging neutron stars (NS) and/or black holes (BH) from the centers of their host galaxies,
as predicted by their formation scenario combined with motion in galactic potentials. 
Furthermore, we estimate the formation rate and merging rate of these massive double degenerate binaries.
The latter is very important for the prospects of detecting gravitational waves with LIGO/VIRGO.
We find that the expected detection rate for LIGO~II is $\sim 850$~yr$^{-1}$ for galactic field sources and that this rate is completely
dominated by merging BHBH binaries. Even LIGO~I may detect such an event ($\sim 0.25$~yr$^{-1}$).
Our preferred model estimate the Galactic field NSNS merger rate to be $\sim 1.5\times 10^{-6}$~yr$^{-1}$.
For BHBH systems this model predicts a merger rate of $\sim 9.7\times 10^{-6}$~yr$^{-1}$.
Our studies also reveal an accumulating numerous population of very wide orbit BHBH systems which never merge ($\tau \gg \tau _{\rm Hubble}$).
\end{abstract}



\section{Introduction}

   It has been recognized for a long time that the time duration of GRBs is bimodal: the majority
(75\%) of the bursts have a long duration with a mean of $\sim 20$~sec., the rest have a much
shorter duration with a mean of only $\sim 0.2$~sec. (Mazets~et~al. 1981 and Hurley~et~al. 1992).
Both of these two classes of GRBs have similar isotropic spatial distributions, but differ with respect to
spectral hardness, fluence and temporal pulse properties (e.g. Kouveliotou~et~al. 1993; Lee \& Petrosian 1997
and Norris~et al. 2000). It is natural to suggest that two distinct types of progenitors are responsible for
the existence of the two observed distinct populations of GRBs.

Since it has been established from measurements
of GRB redshifts that their distance scales are cosmological (Metzger~et~al. 1997), GRB scenarios require
an energy release output of roughly $(\Omega _\gamma /4\pi)\times 10^{54}$~erg, where $\Omega _\gamma$ is the beaming angle. 
According to the internal shock model 
for the production of the observed $\gamma$-rays (e.g. Piran 2000, and references therein), the duration
of a burst is very likely to be a direct measure of the time interval during which the powering
engine is active. In both of the main scenarios the basic ingredient is a black hole surrounded by 
an accretion disk of
nuclear matter threaded by a strong magnetic field. A large disk viscosity causes the matter to spiral inwards
on a timescale of minutes. The poloidal part of the magnetic field in the inner disk, spinning
at millisecond periods, will then accelerate a small amount of material (less than $10^{-5}\,M_{\odot}$) in
the form of a narrow and highly relativistic jet (this is the so-called Blandford-Znajek mechanism,
e.g. Lee, Wijers \& Brown 2000).
Alternatively, the jet of material is produced by annihilating neutrinos along the disk axis. These neutrinos
result from the release of binding energy of the rapidly accreted ($>0.01\,M_{\odot}$) material. 
In either mechanism, the relativistic jet becomes a fireball and ultimately produces the GRB.

The most promising models to account for the
long and short duration bursts involve
the collapse of a massive star (e.g. Woosley~1993 and MacFadyen \& Woosley 1999) and the merging of compact objects
caused by the in-spiral due to gravitational wave emission
in a tight binary (e.g. Goodman 1986; Eichler~et~al. 1989; Paczy\'nski 1990 and Meszaros \& Rees 1992), respectively.
These associations are supported by simulations of outburst durations which are estimated to be 
$\sim \!$ several tens of seconds for the relativistic outflow generated by the collapse of a massive star
(MacFadyen \& Woosley 1999), and $< 1$~sec (Lee \& Kluzniak 1998; Ruffert \& Janka~1999) for the neutrino-driven wind
of merging compact objects, respectively. It has been demonstrated (Meszaros, Rees \& Wijers 1998) that both models
are able to produce the required GRB energies. 

An important method to test the above scenarios is by determining the locations and environments in which
the GRBs must occur according to their respective models. Collapsars (collapsing massive stars) are
expected to be found close to their place of birth as a result of the short lifetime (a few Myr)
of their progenitors. Therefore, these GRBs should occur in dense, dusty star-forming regions -- i.e.
inside their host galaxies. This hypothesis is in agreement with the observations of long-duration GRB afterglows:
there seems to be a correlation between GRB locations and the UV light of their host galaxies
(Sahu~et~al. 1997; Fruchter~et~al. 1998; Kulkarni~et~al. 1998, 1999; Bloom, Kulkarni \& Djorgovski 2002). Furthermore, an underlying
supernova component (a ``bump'') seems to be present in the light curves of GRB~980326 (Bloom~et~al. 1999) 
and GRB~970228 (Reichart 1999; Galama~et al. 2000), and recent detections of an iron line in the
afterglow of several GRBs also provides evidence for the presence of a dense matter in the vicinity
of the bursts (e.g. Yoshida~et~al. 1999; Piro~et~al. 1999, 2000; Vietri~et~al. 1999). 

All of the above mentioned afterglow observations only originate from long-duration GRBs. So far, no similar types
of observations have been possible for the short bursts. It is the hope that observations of the short
bursts will be possible in the near future. An analysis of the expected afterglow properties of these bursts
is, for example, given by Perna \& Belczynski (2002), while the prospect of future observations of these
afterglows are discussed by Panaitescu, Kumar \& Narayan (2001). The main topic to be investigated is the distribution
of merging compact binaries with respect to their host galaxies. The reason is that the two supernova explosions
in the progenitor binaries result in systemic recoil velocities of typically a few hundred~km~s$^{-1}$, and therefore the
merging compact binaries have the potential to travel far from their birth places -- if their merger timescale is 
relatively large. However, the latter issue depends on the final orbital separation of the binaries, after the
second star collapses to form a supernova. 

To estimate the merger timescales (and thus the offset of the associated GRBs
from their birth places) it is necessary to perform a binary population synthesis study of their progenitors. 
Such studies have been performed recently in a number of papers: e.g. Bloom, Sigurdsson \& Pols (1999); 
Belczynski, Kalogera and Bulik (2002); Perna \& Belczynski (2002); Belczynski, Bulik and Kalogera (2002);
Sipior \& Sigurdsson (2002).
The applied formation scenarios, input physics and subsequent results of these studies are not conclusive. 
Our major concern, however, is that all of the above papers have assumed rather simplified calculations 
e.g. for the very important common envelope and spiral-in phase by using a constant value for the 
so-called $\lambda$-parameter. 

Here we use Monte Carlo simulations to determine the range of potential runaway velocities, orbital
separations and merger timescales as predicted by the standard model for the formation of neutron star/black hole
binaries. We include the many recent developments
in the field which severely affects the input physics and the outcome -- especially the helium star evolution
and the binding energy during the common envelope evolution. Also the question of kicks imparted to newborn
black holes has been revisited after the recent detection of the space velocity of the black hole binary
GRO~J1655--40 (Mirabel et~al. 2002).
Our binaries are launched into galaxies with various masses.
Initially the binaries are placed in starforming regions and we keep track of their motion caused by recoil
impacts from the supernova explosions of the stellar components. The resulting merger site distributions of the
compact binaries are then explored for the different galactic potentials.

Population synthesis of massive binaries is also an important tool
to constrain the local merging rate of NS/BH-binaries in order to predict
the number of events detected by the gravitational wave observatories LIGO/VIRGO.  
As a compact binary continue its inspiral, the gravitational waves sweep upward in frequency
from about 10~Hz to $10^3$~Hz, at which point the compact stars will collide and coalesce. It is
the last 15~minutes of inspiral, with $\sim 16\,000$ cycles of waveform oscillation, and the final
coalescence, that LIGO/VIRGO seeks to monitor. LIGO~I and LIGO~II (the advanced interferometer) are expected to detect
NSNS inspiral events out to a distance of $\sim 20$~Mpc and $\sim 300$~Mpc, respectively, according
to recent estimates (Thorne~2001). However, as a result of their larger chirp masses, 
merging double black hole binaries (BHBH) are detected out to
a distance which is $\sim$5~times larger (see e.g. Sipior \& Sigurdsson 2002). 
We have therefore extended our work to also include these systems.

It is the hope that future observations will reveal a combined gravitational wave and GRB source.
This should be possible given the angular resolution of $\sim$1 degree on the sky from simultaneous detections 
using both LIGO and VIRGO. To establish such a connection between these two events would reveal interesting
new insight to the physics involved. Also by
comparing the arrival times of the gravitational waves and the earliest gamma rays,
it should be possible to measure the relative propagation speeds of light and gravitational waves to an accuracy $\sim 10^{-17}$
(Thorne 2001).

In our binary notation we distinguish between BHNS, where the black hole is formed {\it before} the
neutron star, and NSBH where the black hole is formed {\it after} the neutron star. NSNS and BHBH
refer to double neutron star and double black hole systems, respectively. In Section~2 we describe our model
for making neutron star/black hole binaries via interacting binary evolution. Our modelling of galactic potentials
and binary star formation rates are presented in Section~3. In Section~4 we summarize our results and in
Section~5 we follow up with discussions and compare with previous work. Finally, our conclusions are given in Section~6.


\section{Formation of neutron star/black hole binaries}
\begin{figure}
  \centering
  \includegraphics[height=10.0cm]{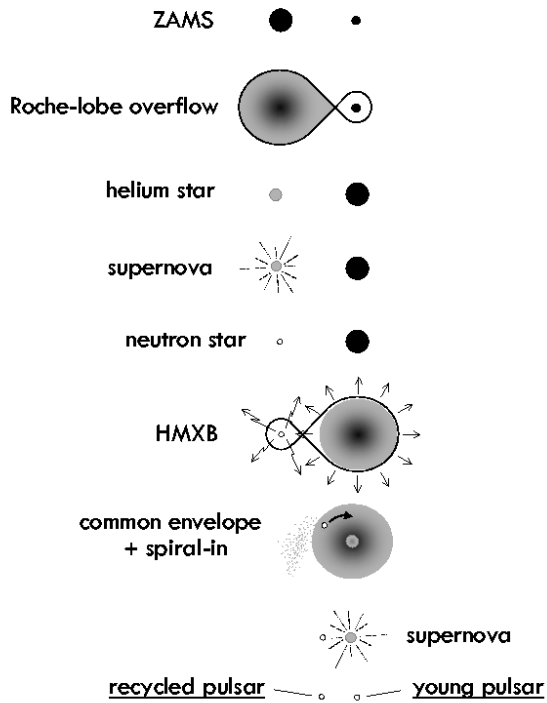}
  \caption{Cartoon depicting the formation of a HMXB and finally
           a double neutron star (and/or black hole binary system). Such a binary will experience two supernova
           explosions. It is always the recycled pulsar which is observed in a double pulsar system as a result of
           its very long spin-down timescale compared to a young pulsar (a factor of $\sim 10^2$).
           If the final system is very tight it will coalesce as a result of gravitational wave radiation.}
  \label{HMXB-cartoon}
\end{figure}
In Fig.~1 we have shown the standard evolutionary sequences leading to the formation of a double neutron star system.
The (recycled) neutron stars are detected as radio pulsars. The scenario for producing a black~hole/neutron~star binary
system is similar. As can be seen from the figure the progenitor systems evolve through a high-mass {X}-ray binary (HMXB) phase.
The formation of a HMXB requires two relatively massive stars ($>12\,M_{\odot}$) on the ZAMS. 
Alternatively, the secondary 
star can be less massive initially, as long as it gains enough material from the primary star so that it will
later end up above the threshold mass for undergoing a supernova explosion (like the primary star).
The first mass transfer phase, from the primary to the secondary star, is usually assumed to be dynamically stable
(semi-conservative) since the
mass ratio, at the onset of the RLO, is not too extreme. However later on, when the secondary star evolves,
{\em all} HMXBs must end up in a common envelope phase,
as the neutron star (or low-mass black hole) is engulfed by the extended envelope of its companion, in an orbit
which is rapidly shrinking due to significant loss of orbital angular momentum.
This stage of binary evolution is extremely important, since the outcome is a huge reduction of the orbital separation
or merging of the stellar components -- see Sect.~\ref{CE}.
Stellar winds of massive stars, as well as of naked helium cores (Wolf-Rayet stars), are also some of the most uncertain
aspects of the modelling of HMXB evolution. Finally, the physical conditions which determine the formation of
a neutron star versus a black hole are also still quite unknown. 
Below we shall briefly outline our model of binary evolution. 

\subsection{Stellar evolution models}
\begin{figure}
   \centering
   \includegraphics[height=7cm]{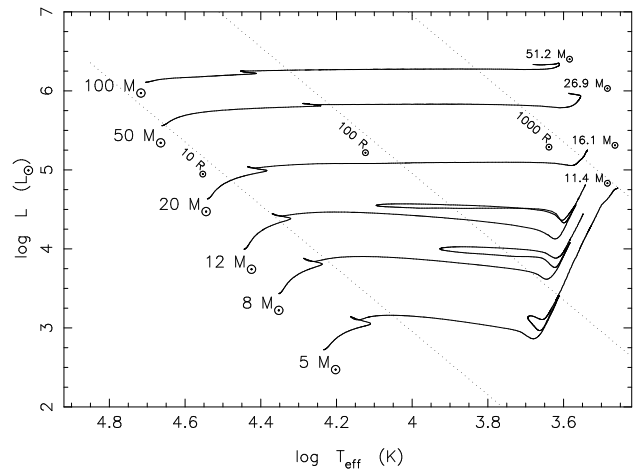}
      \caption{Evolutionary tracks for a few selected stars in our code ({X}=0.70, {Z}=0.02). For the massive stars
               ($12-100\,M_{\odot}$) the final mass, after stellar wind mass loss, is written at the end of the tracks.}
         \label{HR}
   \end{figure}

We used Eggletons numerical stellar evolution code to collect a large array of grid points
for stars with masses in the interval $0.9-100\,M_{\odot}$. 
This code uses a self-adaptive, non-Lagrangian mesh-spacing
which is a function of local pressure, temperature, Lagrangian mass and radius.
It treats both convective and semi-convective mixing as a diffusion process and
finds a simultaneous and implicit solution of both the stellar structure
equations and the diffusion equations for the chemical composition. New
improvements are the inclusion of pressure ionization and Coulomb interactions
in the equation-of-state, and the incorporation of recent opacity tables,
nuclear reaction rates and neutrino loss rates. The most important recent
updates of this code are described in Pols~et~al. (1995, 1998) and some are
explained in Han~et~al. (1994). In our standard model we used a chemical composition of
({X}=0.70, {Z}=0.02) and assumed a mixing-length parameter of $\alpha = l/H_p =2.0$.
The uncertainty of these values (and their temporal evolution) is not very significant for
the results of our work in consideration here, given the many major uncertainties of
parameters governing the mass-transfer phases, spiral-in, helium star wind, supernova explosion etc. 
Convective overshooting is taken into account in the same way as in Pols~et~al. (1998)
using an overshooting constant $\delta _{\rm ov} = 0.10$. We used de~Jager's work 
(de~Jager~et~al. 1988; Nieuwenhuijzen \& de~Jager 1990) to estimate the wind mass loss prior to
the mass-transfer phase. Examples of our stellar evolutionary tracks are shown in Fig.~\ref{HR}. 
We calculated $\sim \!40\,000$ grid points of fine mesh spacing in the HR-diagram in order 
to perform fast Monte Carlo population synthesis.

\subsection{A model of binary evolution}
In general terms our model follows the standard model
for the formation and evolution of HMXBs. Therefore, in the following we shall mainly discuss the issues below where our model
applies different input physics compared to earlier work. For a general discussion of close binary evolution
we refer to, for example, van~den~Heuvel~(1994) and Tauris \& van~den~Heuvel~(2003).

To simulate the formation of massive double degenerate binaries we assume that the initial binary system
consists of two zero-age main sequence (ZAMS) stars, and chose primary masses in the interval
$10-100\,M_{\odot}$ using a Salpeter-like IMF, $N(m)\propto m^{-2.70}$ (Scalo 1986). The secondary stars (less massive
than their primary companions) were chosen in the interval $4-100\,M_{\odot}$ according to
the mass-ratio function: $f(q)=2/(1+q)^2$ (Kuiper 1935). We adopt the term ``primary'' to refer to
the {\em initially} more massive star, regardless of the effects of mass transfer or loss as the system evolves.
The initial orbital separations were chosen between $5-10\,000\,R_{\odot}$ from a distribution flat in $\log a$.

\subsection{Helium stars, SN kicks and remnant masses}
\begin{figure}
   \centering
   \includegraphics[height=7cm]{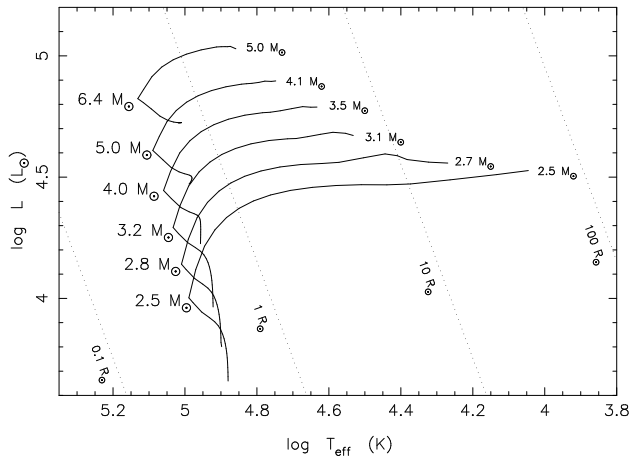}
      \caption{Evolutionary tracks for selected helium stars in our code ({Y}=0.98, {Z}=0.02). The final mass 
               (after stellar wind mass-loss) is written at the end of the tracks. 
               The expansion of these (low-mass) helium stars in close binaries often results in a secondary 
               mass-transfer phase.
               The more massive Wolf-Rayet stars do not expand very much during their evolution.
               Almost all the {He} is burned into {C}, {O} and subsequently {Ne}.
               After O.~Pols (priv. comm.)}
         \label{HRHe}
   \end{figure}

For low-mass stars, the mass of the helium core in post main-sequence evolution is practically
independent of the presence of an extended hydrogen-rich envelope. However, for more massive
stars ($> 2\,M_{\odot}$) the evolution of the core of an isolated star differs from that of a
naked helium star (i.e. a star which has lost
its hydrogen envelope via mass transfer in a close binary system).
It is very important to incorporate the giant phases of helium star evolution.
Of particular interest are the low-mass helium stars ($M_{\rm He} < 3.5\,M_{\odot}$) since they
swell up to large radii during their late evolution (e.g. Habets~1986).
This may cause an additional phase of mass transfer from the naked helium star to its companion
(often referred to as case~BB mass transfer). Recent detailed studies of helium stars
in close binaries have been performed by Dewi~et~al. (2002), Dewi \& Pols (2003) and Ivanova~et~al. (2003). 
It should be noticed that helium cores in binaries have tiny envelopes of hydrogen ($<0.01\,M_{\odot}$)
when they detach from Roche-lobe overflow (RLO) mass transfer. 
This tiny envelope seems to have important effects on the subsequent radial evolution of the helium star (e.g. Han~et~al.~2002).\\
The evolution of more massive helium stars (Wolf-Rayet stars) is also quite important. There is currently
not a clear agreement on the rate of intense wind mass-loss from Wolf-Rayet stars (e.g. Wellstein \& Langer~1999;
Nugis \& Lamers~2000; Nelemans \& van~den~Heuvel~2001).
A best estimate fit to the wind mass-loss rate of Wolf-Rayet stars is, for example, given by Dewi~et~al.~(2002).
We have adopted their helium star models ({Z}=0.02, {Y}=0.98), including wind mass-loss, 
as calculated by Onno Pols (private communication) -- see Fig.~\ref{HRHe}.  
The uncertainty in determining the wind mass-loss rate also affects the threshold mass for core collapse leading to a black hole
(Schaller~et~al. 1992; Woosley, Langer \& Weaver 1995; Brown, Lee \& Bethe 1999).
However, it may well be that core mass is
not the only important factor to determine the outcome. Magnetic field and spin of the collapsing core
could also play a major role (Ergma \& ~van~den~Heuvel 1998; Fryer \& Heger 2000). Furthermore, it seems clear from observations
that there is an overlap in the mass range for making a neutron star versus a black hole.
In our code we adopted the threshold values given in Table~\ref{Remnant}. The threshold mass for the {CO}-core corresponds
to a helium star which has experienced another RLO (case~BB) in its giant phase and thereby lost its helium envelope.
   \begin{table}
      \caption[]{Threshold masses for producing black holes and neutron stars}
         \label{Remnant}
         \begin{center}
         \begin{tabular}{cccc}
            \hline
            \noalign{\smallskip}
            Remnant & Helium core & {CO}-core & $M_{\rm ZAMS}^{\rm min}$\\
            \noalign{\smallskip}
            \hline
            \noalign{\smallskip}
            Black hole & $5.5\,M_{\odot}$ & $4.0\,M_{\odot}$ &   $22\sim 25\,M_{\odot}\,^{*}$ \\
            Neutron star & $2.8\,M_{\odot}$ & $1.7\,M_{\odot}$ & $10\sim 12\,M_{\odot}\,^{*}$  \\
            \noalign{\smallskip}
            \hline
         \end{tabular}
\begin{list}{}{}
\item[$^{\mathrm{*}}$] Depending on the evolutionary status at the onset of RLO (case~{C} or {B}).
\end{list}
         \end{center}
   \end{table}

\begin{figure}
   \centering
   \includegraphics[height=7cm]{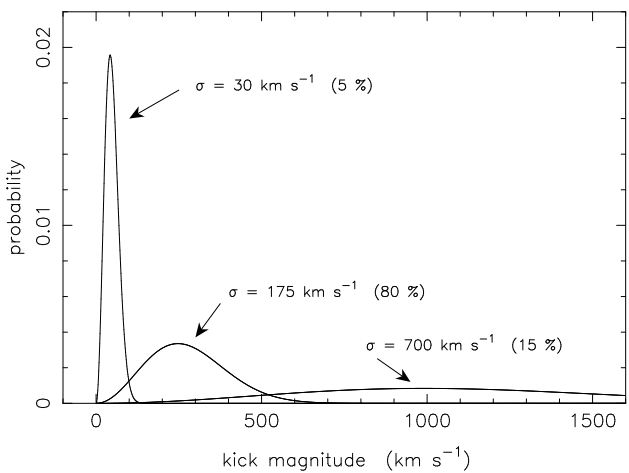}
      \caption{A momentum kick is assumed to be imparted to all newborn neutron stars.
               The kick magnitude of the asymmetric SN was drawn from a weighted
               sum of the three Maxwellian distributions shown above. 
               The direction of the kicks is assumed to be isotropic.
               Black holes were assumed to be accompanied with a smaller kick velocity -- see text
               for description.} 
         \label{kick}
   \end{figure}

We assumed all neutron stars to be formed in asymmetric SNe with a mass of $1.3\,M_{\odot}$ and
chose the 3-D magnitude of the kicks from a weighted sum of three Maxwellian distributions of speeds
with $\sigma _w = 30, 175$ and 700~km~s$^{-1}$ (5\%, 80\% and 15\%, respectively -- see Fig.~\ref{kick}).
The choice of this distribution is partly motivated by the study of Cordes \& Chernoff (1998), but
also includes a low-velocity component to account for the many neutron stars retained in rich
globular clusters like 47~Tuc (e.g. Pfahl~et~al. 2002).
Furthermore, we assumed the direction of the kicks to be isotropic (see however Lai, Chernoff \& Cordes~2001) and
neglected any impact on the companion star from the ejected shell. 
The recent determination of the run-away velocity ($112\pm18$~km~s$^{-1}$) of the black hole binary GRO~J1655--40 
(Mirabel et~al. 2002) seems to suggest that also a stellar collapse leading to the formation of a black hole 
is accompanied by a momentum kick. In our standard model we apply a reduced kick in the formation process
of black holes. We do this by assuming that a similar linear momentum, comparable to that imparted to neutron stars, is 
given to black holes, such that their resulting kick velocities are reduced by a factor of $M_{\rm BH}/M_{\rm NS}$
with respect to our kick distribution applied on newborn neutron stars. For the masses of the black holes 
we assumed a mass loss fraction of 30\% during the stellar collapse (Nelemans, Tauris \& van~den~Heuvel 1999).

\subsection{Common envelope and spiral-in evolution}
\label{CE}
Except for a few cases in which the two stars in a massive binary are born on the ZAMS with
almost equal masses (Brown~1995), the system will evolve through a HMXB-phase and subsequently enter
a common envelope and spiral-in phase. In our code we follow the treatment of common envelope evolution
introduced by Dewi \& Tauris (2000). The same method has been applied recently by others, e.g. 
Podsiadlowski, Rappaport \& Han~(2002).
Hence, we calculate the binding energy of the stellar envelope 
by integrating through the outer layers of the donor star:
\begin{eqnarray}
         E_\mathrm{bind} & = & - \int_{M_\mathrm{core}}^{M_\mathrm{donor}}
         \frac {G M(r)} {r} dm + \alpha_\mathrm{th}
         \int_{M_\mathrm{core}}^{M_\mathrm{donor}} U dm
         \label{envhan}
        \end{eqnarray}
where the first term is the gravitational binding energy and $U$ is the internal
thermodynamic energy (Han~et~al.~1994, 1995).
The value of $\alpha_\mathrm{th}$ depends on the details of the ejection
process, which are very uncertain. A value of $\alpha_\mathrm{th}$ equal to 0 or
1 corresponds to maximum and minimum envelope binding energy, respectively.
By adopting the energy formalism of Webbink (1984) and de~Kool (1990), we then equate this envelope binding energy
with the release of orbital energy in order to estimate the reduction of the orbit during spiral-in:
$E_{\rm bind} \equiv \eta_{\rm CE} \,\, \Delta E_{\rm orb}$
where $\eta_\mathrm{CE}$ describes the efficiency of ejecting the envelope, i.e.
of converting orbital energy into the kinetic energy that provides the outward
motion of the envelope.
The total change in orbital energy is then
simply given by: 
        \begin{eqnarray}
         \Delta E_\mathrm{orb} & = &
         - \frac {G M_\mathrm{core} M_\mathrm{1}} {2 \,a_\mathrm{f}}
         + \frac {G M_\mathrm{donor} M_\mathrm{1}} {2 \,a_\mathrm{i}}
         \label{lambda:eq:orb}
        \end{eqnarray}
where $M_\mathrm{core} = M_\mathrm{donor} - M_\mathrm{env}$ is the mass of the
helium core of the evolved donor star; $M_\mathrm{1}$ is the mass of the
companion star; $a_\mathrm{i}$ is the initial separation at the onset of the RLO
and $a_\mathrm{f}$ is the final orbital separation after the CE-phase.
The orbital separation  of the
surviving binaries is quite often reduced by a factor of $\sim\!100$ as a result of the spiral-in.
If there is not enough orbital energy available to eject the envelope the stellar components will coalesce.
For massive stars ($M>10\,M_{\odot}$) this method results in so-called $\lambda$-values $< 0.1-0.01$ 
(Dewi \& Tauris~2001; Podsiadlowski, Rappaport \& Han~2002).
These values are much smaller than the constant value of $\lambda \sim 0.5$ often used in the literature 
(e.g. Hurley, Tout \& Pols 2002; Sigurdsson \& Pols 1999; Fryer, Woosley \& Hartmann 1999;
Portegies~Zwart \& Yungelson 1998, or alternatively, $\lambda \,\eta _{\rm CE}$ is used as a single free parameter: 
Belczynski, Kalogera \& Bulik 2002; Belczynski, Bulik \& Kalogera 2002; Perna \& Belczynski 2002).
This result has the very important consequence that in our study many HMXBs will
not produce double neutron star/black hole systems because they merge during their subsequent CE-phase
(however, those systems that do survive are generally in tight orbits often leading to additional helium star mass transfer 
and formation of double degenerate systems merging on a short timescale).\\
   \begin{figure}
   \centering
   \includegraphics[height=7cm]{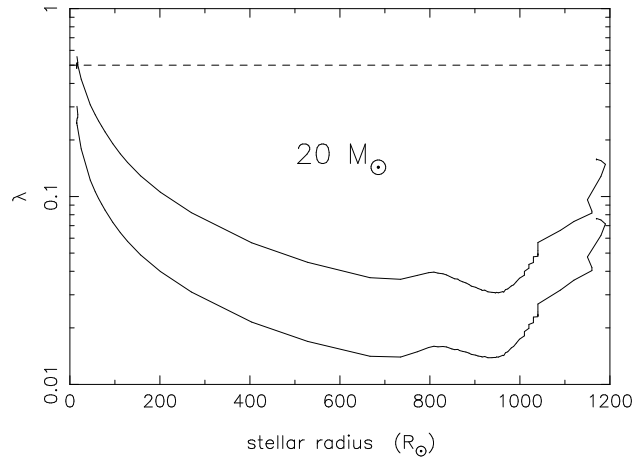}
      \caption{The $\lambda$-parameter for a $20\,M_{\odot}$ star as a function of stellar radius. The upper curve includes internal thermodynamic
               energy ($\alpha _{\rm th}=1$) whereas the lower curve is based on the sole gravitational binding energy ($\alpha _{\rm th}=0$) 
               -- see Eq.~\ref{envhan}. There is a factor $\sim$2 in difference between the $\lambda$-curves in
               accordance with the virial theorem. It is a common misconception to use a constant (and large) value
               of $\lambda \simeq 0.5$ marked by the dashed line. In this plot the stellar core is defined as the region
               where {X}$<0.10$. See text.}
         \label{lambdafig}
   \end{figure}
It should be noticed that the exact determination of $\lambda$ depends on how the core boundary is defined
(see Tauris \& Dewi 2001 for a discussion). For example, if the core boundary (bifurcation point of envelope ejection in a CE)
of the $20\,M_{\odot}$ in Fig.\ref{lambdafig} is moved out by $0.1\,M_{\odot}$ then $\lambda$ is typically increased 
by a factor of $\sim\! 2$.

\subsubsection{Helium star common envelope}
We take into account that some helium stars, in their giant stages, regain contact with their compact companion star.
This leads to a second mass-transfer phase resulting in a naked {CO}-core, if the stellar components do not coalesce.
From detailed helium star analysis, Dewi \& Pols (2003) discovered that there is a critical mass 
($M_{\rm He}^{\rm crit}\sim 3.3\,M_{\odot}$)
above which the RLO from the helium star to a neutron star is dynamically stable. Below this value the system will evolve into
a CE and spiral-in phase as a consequence of the rapid expansion of the low-mass helium star. We adopted this value in our code
to determine the fate of helium star CE (see however Ivanova~et~al. 2003 for a different point of view). 
Typical helium star $\lambda$-values are $0.01-0.2$ depending on radius and $\alpha _{\rm th}$. 

\subsubsection{Released accretion energy during CE evolution}
Even though the spiral-in timescale is very short ($<1000$ yr) for common envelope evolution (Taam \& Sandquist 2000),
and the accretion rate is limited by the photon radiation pressure (the Eddington limit), the gravitational
potential energy release from the $\sim 10^{-5}\,M_{\odot}$ accreted onto the neutron star or black hole
contribute to expel the envelope of the donor star.
We modelled this effect by assuming a total energy release of 
$\Delta E_{\rm acc} \sim \tau _{\rm ce}\,L_{\rm Edd} = 3.8\times 10^{48} \,(M_{\rm X}/M_{\odot})$~erg,
and twice this amount for helium star CE, where $M_{\rm X}$ is the mass of the compact object. 
Given the high column density inside the common envelope a significant part of this energy will be absorbed and facilitate
the ejection of the envelope and thereby increase the post-CE orbital separation by a factor:
\begin{equation}
  \label{Eacc}
  a_{\rm f}^{*} = a_{\rm f}\;(1+\Delta E_{\rm acc}/E_{\rm bind}) 
\end{equation}
The average value of this factor is $\sim \!1.8$ for all our systems evolving through a CE-phase (the typical value is
only $\sim \!1.3$, but a few systems with very evolved donors have $\Delta E_{\rm acc}>E_{\rm bind}$ and a corresponding 
correction factor up to about~5).

\subsection{RLO and the orbital angular balance equation}
The orbital angular momentum of a binary system is given by:
\begin{equation}
  J_{\rm orb} = |\vec{r}\times\vec{p}| = \frac{M_1 M_2}{M}\,\Omega \,a^2 \,\sqrt{1-e^2}
~\label{Jorb}
\end{equation}
where $a$ is the separation between the stellar components, $M_1$ and $M_2$ are the masses of
the accretor and donor star, respectively, $M=M_1 + M_2$ and the orbital angular velocity,
$\Omega = \sqrt{GM/a^3}$. Here $G$ is the constant of gravity.
Tidal effects acting on a near-RLO (giant) star will circularize the
orbit on a short timescale of $\sim 10^4$~yr (Verbunt \& Phinney 1995) and
in the equation below we therefore disregard any small eccentricity ($e=0$).
A simple logarithmic differentiation of the above equation yields the rate of change in orbital separation:
\begin{equation}
  \frac{\dot{a}}{a} = 2\,\frac{\dot{J}_{\rm orb}}{J_{\rm orb}} - 2\,\frac{\dot{M}_1}{M_1}
                    - 2\,\frac{\dot{M}_2}{M_2} + \frac{\dot{M}_1 + \dot{M}_2}{M}
  \label{adot}
\end{equation}
where the total change in orbital angular momentum is:
\begin{equation}
  \frac{\dot{J}_{\rm orb}}{J_{\rm orb}} = \frac{\dot{J}_{\rm gwr}}{J_{\rm orb}}
       +\frac{\dot{J}_{\rm mb}}{J_{\rm orb}} + \frac{\dot{J}_{\rm ls}}{J_{\rm orb}}
       +\frac{\dot{J}_{\rm ml}}{J_{\rm orb}}
  \label{Jdot}
\end{equation}
These two equations constitute the orbital angular momentum balance equation. The first term on the right-hand side of
this equation gives the change in orbital angular momentum
due to gravitational wave radiation (Landau \& Lifshitz~1958).
The second term in Eq.~\ref{Jdot} arises from so-called magnetic braking (this mechanism is not relevant
for HMXBs).
The third term ($ \dot{J}_{\rm ls}/J_{\rm orb}$) on the right-hand side of Eq.~(\ref{Jdot})
describes possible exchange of angular momentum
between the orbit and the donor star due to its expansion or contraction (Tauris~2001).
Finally, the last term in Eq.~(\ref{Jdot}) represents the change in orbital angular momentum
caused by mass loss from the binary system. This is usually the dominant term in the orbital angular momentum
balance equation and its total effect is given by:
\begin{equation}
  \frac{\dot{J}_{\rm ml}}{J_{\rm orb}} = \frac{\alpha + \beta q^2 + \delta\gamma(1+q)^2}{1+q}\,
                                         \frac{\dot{M}_2}{M_2}
\end{equation}
where $q=M_2/M_1$ is the mass ratio, and 
where $\alpha$, $\beta$ and $\delta$ are the fractions of mass lost from the donor in the form
of a direct fast wind, mass ejected from the vicinity of the accretor and from a circumbinary
coplanar toroid (with radius, $a_r = \gamma ^2 a$), respectively -- see van~den~Heuvel~(1994) and
Soberman, Phinney \&  van~den~Heuvel~(1997). The accretion efficiency of the accreting star is thus
given by: $\epsilon = 1 -\alpha -\beta -\delta$, or equivalently:
\begin{equation}
  \partial M_1 = -(1 -\alpha -\beta -\delta)\, \partial M_2
\label{epsilon}
\end{equation}
where $\partial M_2 < 0$ ($M_2$ refers to the donor star).
These factors will be functions of time as the binary system evolves during the mass-transfer phase.
The general solution for calculating the change in orbital separation during the {X}-ray phase
is found by integration of the orbital angular momentum balance equation (Eq.~\ref{adot}):
\begin{eqnarray}
  \frac{a}{a_0} & = & \Gamma _{ls}\,\displaystyle \left( \frac{q}{q_0}\right) ^{2\,(\alpha + \gamma\delta -1)}\;
                \left(\frac{q+1}{q_0+1}\right) ^{\textstyle\frac{-\alpha - \beta +\delta}{1-\epsilon}}\\ \nonumber
             & &  \times \; \left(\frac{\epsilon q+1}{\epsilon q_0+1}\right) ^{3+2\,\textstyle\frac{\alpha \epsilon ^2 + \beta +
                      \gamma\delta (1-\epsilon )^2} {\epsilon (1-\epsilon)}}\\ \nonumber
\label{aa0}
\end{eqnarray}
where the subscript `0' denotes initial values and $\Gamma _{ls}$ is factor of order unity
to account for tidal spin-orbit couplings ($\dot{J}_{\rm ls}$).

In our standard model for the RLO we assumed a direct wind mass-loss fraction of $\alpha =0.20$ and neglected the possibility
of any circumbinary toroid (i.e. $\delta = 0$). Furthermore, we assumed the accretion rate onto the compact object to
be limited by the Eddington luminosity and hence: 
$\beta \equiv max \left( (|\dot{M}_{\rm donor}| - \dot{M}_{\rm Edd})/|\dot{M}_{\rm donor}|,0\right)$,
where $|\dot{M}_{\rm donor}|$ is the mass-transfer rate from the donor star. For a typical neutron star accreting hydrogen
$\dot{M}_{\rm Edd} \simeq 1.5\times 10^{-8}\,M_{\odot}$~yr$^{-1}$.

\subsection{Gravitational waves and merging NS/BH binaries}
In the Newtonian/quadrupole approximation ($a\ll \lambda _{\rm gwr}$) the merging timescale of a
compact binary with semi-major axis, $a$ and eccentricity, $e$ is given by Peters~(1964):
\begin{equation}
\tau_{\rm gwr} (a_0,e_0)=\frac{12}{19}\frac{C_0^{4}}{\beta}\times \int_{0}^{e_0}\frac{e^{29/19}[1+(121/304)e^{2}]^
{1181/2299}}{(1-e^{2})^{3/2}}de
\end{equation}
where
\begin{equation}
C_0=\frac{a(1-e^{2})}{e^{12/19}}\,[1+(121/304)e^{2}]^{-870/2299}
\end{equation}
can be found from the initial values ($a_0, e_0$), and the constant $\beta$ is given by:
\begin{equation}
\beta =\frac{64G^3}{5c^5}M^2\mu
\end{equation}
Here $M$ and $\mu$ denote the total mass and the reduced mass of the system, respectively.
The integral given above must be evaluated numerically.
However, for circular orbits the merging timescale can easily be found analytically: $\tau _{\rm gwr}^{\rm circ} = a_0^4/4\beta$.
The timescale is very dependent on both $a$ and $e$. Tight and/or eccentric orbits
spiral-in much faster than wider and more circular orbits -- see Fig.~\ref{ae}.
For example, we find that the double neutron star system PSR~1913+16
($P_{\rm orb}=7.75$~hr, $M_{\rm NS}=1.441$ and $1.387\,M_{\odot}$, respectively)
with an eccentricity of 0.617 will merge in 302 Myr; if its orbit
was circular the merger time would be five times longer: 1.65~Gyr!

\section{Star formation rates, galactic models and starforming regions}
We have assumed that the star formation rate has been continuous in the Galactic disk
as inferred from observations (e.g. Gilmore~2001).
Furthermore, following Hurley, Tout \& Pols~(2002) we assume that one binary ($M_1 > 0.8\,M_{\odot}$) is born in the Galaxy per year.
Hence, for the formation rate of a massive binary with primary star mass $> 10\,M_{\odot}$ and secondary
star mass $> 4\,M_{\odot}$ we simply use:
\begin{equation}
  \Phi _{\rm BSFR} = f\,0.01\quad {\rm yr}^{-1}
\label{BSFR}
\end{equation}
where $f$ is a factor $\sim$1 to scale all our derived formation and merger rates.
We fix this rate over the lifetime of the Galaxy ($\sim 12$~Gyr). This prescription should
facilitate comparison with other studies.
 
\subsection{Galactic potentials}
We have evolved our binaries in the potential of spiral galaxies. The
adopted potential is a model for the Milky~Way made by Flynn, Sommer-Larsen \& Christensen (1996) with
a few corrections to the parameters (Sommer-Larsen, private communication
2002). The potential $\Phi (R,z)$ is modelled in cylindrical coordinates: 
$R$ being the distance in the plane from the center of the galaxy, and 
$z$ being the height above the plane. The potential is a sum of a
dark matter halo $\Phi_{H}$, a central component $\Phi_{C}$ and a disk
$\Phi_{D}$. 
The dark matter halo is assumed to be spherically symmetric:
\begin{equation}
\Phi_{H} =\frac{1}{2}V_{H}^{2}\ln (r^2+r_0^2)
\end{equation}
where $r^2=R^2+z^2$.
The central part of the potential is made-up of two spherical components:
\begin{equation}
\Phi_C=-\frac{GM_{C_1}}{\sqrt{r^2+r_{C_1}^2}}-\frac{GM_{C_2}}{\sqrt{r^2+r_{C_2}^2}}
\end{equation}
Here $G$ is the gravitational constant, $M_{C_1}$ and $r_{C_1}$ are the
mass and the core radius of the bulge/stellar halo term, while $M_{C_2}$ and $r_{C_2}$ are the mass and core radius of the inner core.
The disk potential is modelled using a combination of three Miyamoto-Nagai
potentials, see Miyamoto \& Nagai (1975):
\begin{equation}
\Phi_{D_n}=\frac{-GM_{D_n}}{\sqrt{(R^2+[a_n+\sqrt{z^2+b^2}]^2)}}\,,\quad n=1,2,3
\end{equation}
The parameters $b$ and $a_n$ are related respectively to the scaleheight
and scalelength of the disk, and $M_{D_n}$ are the masses of the three
disk components. The parameter values for the Milky~Way are found in
Table~\ref{milk}.\\
To obtain potentials for galaxies with other masses than the Milky~Way,
we have scaled the parameters with the square root of the galaxy mass
(assuming that the ratio of the components of the galaxies remains constant).
This gives a constant surface brightness as observed, see e.g. Binney
\& Tremaine (1994).

Star formation takes place in two very distinct physical
environments. Mostly in the extended disks of spiral and
irregular galaxies, but also a significant fraction in the
compact dense gas disks in
the centers of galaxies.
The central star formation happens in starbursts with very high
star formation rates for 0.1--1.0 Gyr (Kennicut 1998),
while star formation in the extended disk is steady or
slowly exponentially decreasing (Bruzual \& Charlot 1993).
We have assumed that the
distribution of birth places follow the density distribution
of the extended disk. The latter can be found by applying Poisson's equation
to the disk potential:
\begin{equation}
\nabla^{2}\Phi_D=4\pi G\rho_D
\end{equation}
Finally, we assume that the binaries are born with a velocity
equal to the local rotational velocity.
\begin{table}
  \caption{Adopted parameters for the Galactic potential}
  \label{milk}
  \begin{center}
  \begin{tabular}{lcr}
  \hline
                Component       & Parameter	& Value\\
  \hline
                & $r_0$     	& 8.5 kpc\\
		& $V_{H}$	& 210 km~s$^{-1}$\\
Bulge/Stellar halo &$r_{C_{1}}$ & 2.7 kpc\\
		&$M_{C_{1}}$	& 3.0$\times 10^{9}\,M_{\odot}$\\
Central Component&$r_{C_{2}}$	& 0.42 kpc\\
		&$M_{C_{2}}$	& 1.6$\times 10^{10}\,M_{\odot}$\\
Disk		&$b$		& 0.3 kpc\\
		&$M_{D_{1}}$	& 6.6$\times 10^{10}\,M_{\odot}$\\
		&$a_{1}$	& 5.81 kpc\\
		&$M_{D_{2}}$	& -2.9$\times 10^{10}\,M_{\odot}$\\
		&$a_{2}$	& 17.43 kpc\\
		&$M_{D_{3}}$	& 3.3$\times 10^{9}\,M_{\odot}$\\
		&$a_{3}$	& 34.86 kpc\\
  \hline
  \end{tabular}
  \end{center}
\end{table}

\section{Results}
The results presented in this paper
are based on a typical simulation of 20~million ZAMS binaries
with standard parameters as given in Table~\ref{stdpar}. Of these systems 99.9$\,$\% did
not make it to the final stage. Most of these systems either merged during the CE and spiral-in phase
or were disrupted by a supernova explosion.

   \begin{figure}
   \centering
   \includegraphics[height=7cm]{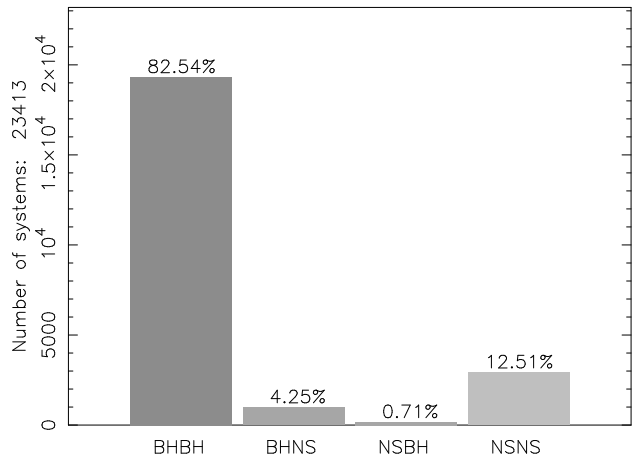}
      \caption{Relative formation ratios of binaries with neutron stars and/or black holes
               which will coalesce within 10~Gyr.}
       \label{histo}
   \end{figure}

\subsection{Relative formation ratios and merging timescales}
\begin{figure*}
   \centering
   \includegraphics[height=10.5cm]{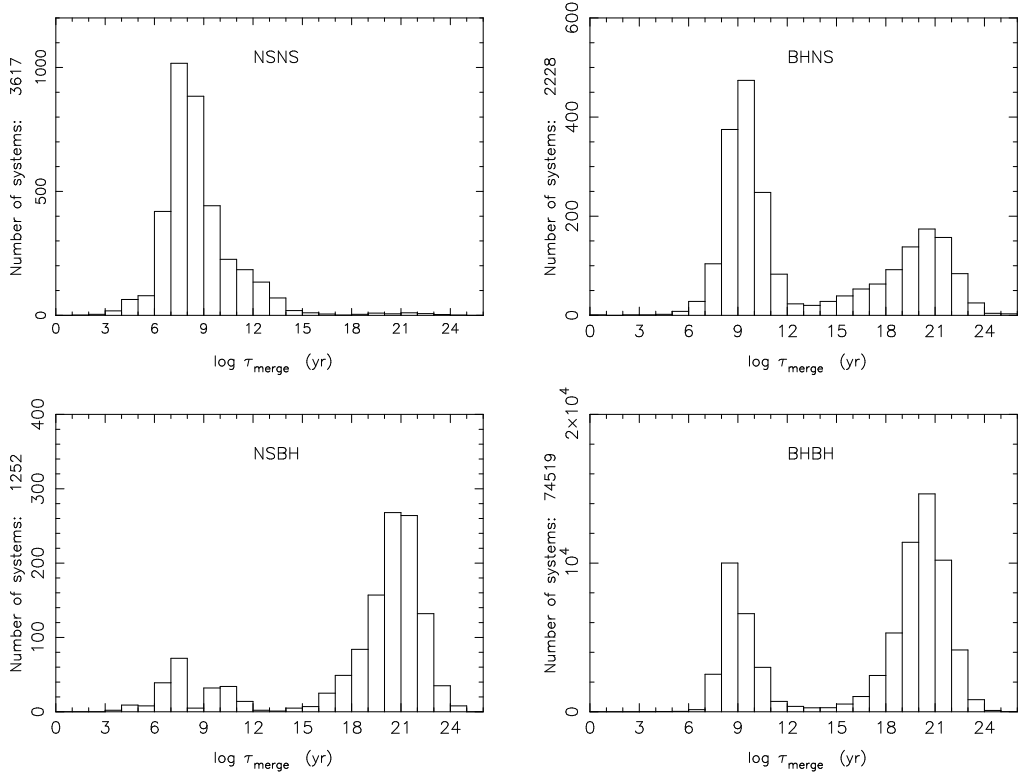}
      \caption{Histograms of merging timescales for binaries with neutron stars and/or black holes.
               The populations of NSBH, BHNS and BHBH are seen to be bimodal with a large component
               of very wide binaries with $\tau _{\rm merge} \sim 10^{21}$~yr. See text for a detailed discussion.}
         \label{merge_hist}
   \end{figure*}

\begin{figure*}
   \centering
   \includegraphics[height=10.5cm]{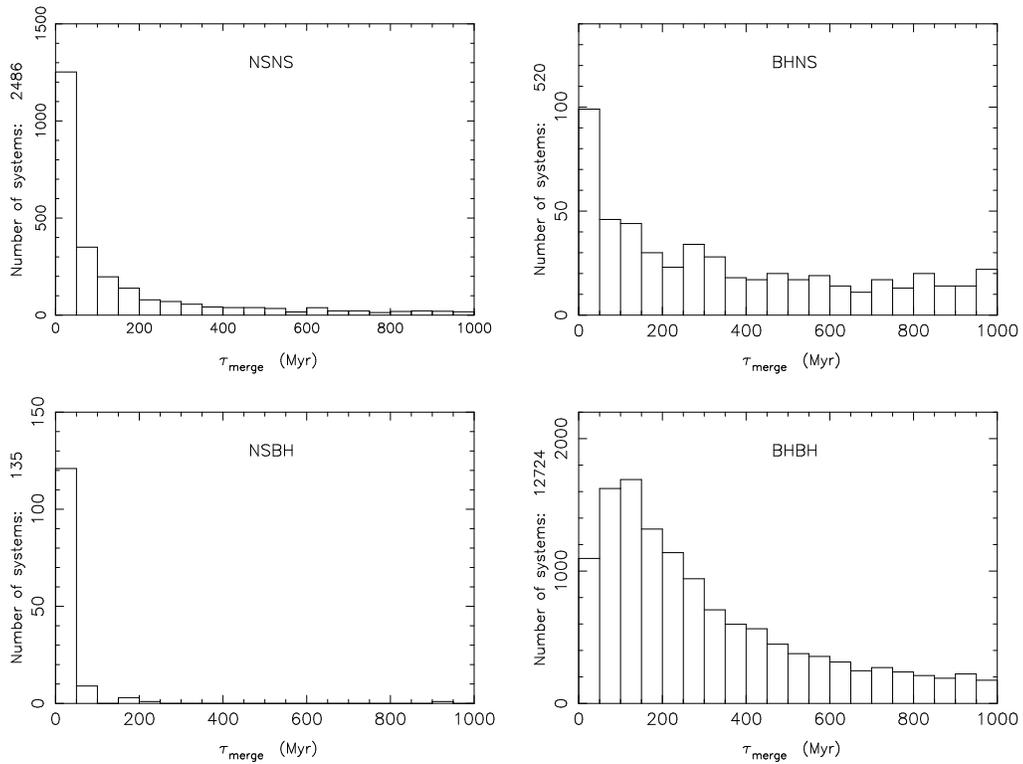}
      \caption{Histograms of merging timescales for binaries with neutron stars and/or black holes which coalesce within 1~Gyr.}
         \label{merge_hist1}
   \end{figure*}

The aim of this paper is to focus on GRB progenitors and burst sources of gravitational wave radiation. Therefore, we
start out by showing in Fig.~\ref{histo} the relative formation ratios of BHBH, BHNS, NSBH and NSNS systems which will merge
within 10~Gyr. A total of 23$\,$413 such sources is formed over a time interval of $\sim$2~Gyr from 20~million
massive ZAMS binaries (see Table~\ref{stdpar} for input parameters). It is seen that double black hole systems (BHBH) 
dominate this population and that only few close-orbit NSBH systems are formed.
However, from Fig.~\ref{merge_hist} it is seen that the relative formation ratios of mergers do not represent
the overall relative formation ratios of all massive double degenerate systems. 
Many NSBH, BHNS and BHBH systems are formed in very wide orbit systems
($P_{\rm orb}\ga 100-1000$~yr) which will never merge ($\tau _{\rm merge} \sim 10^{21}$~yr). For these
three categories the formation of such wide-orbit systems is even seen to be dominating. 
Almost all of the binaries formed 
in this right-hand peak of the bimodal $\tau _{\rm merge}$ distribution are caused by binaries
which did not experience RLO from their secondary star (causing the orbits to widen subsequently from stellar wind mass loss
and the effects of the second SN explosion). The majority of the wide-orbit BHNS systems also evolved without
primary star RLO and thus remained detached binaries during their entire evolution. The same condition applies to
$\sim 1/3$ of the wide-orbit BHBH systems, but none of the NSBH systems. The reason for the latter is that 
without RLO the secondary star will never end up with a mass above the threshold limit for making a black hole.
Another interesting feature is that almost half of all the BHNS systems which merge within 10~Gyr 
evolve without primary star RLO. However, in these systems
the black hole is shot into a closer orbit (due to the kick) which later, as a result of nuclear expansion,
causes the secondary star to fill its Roche-lobe. This gives rise to the formation of
a common envelope and spiral-in phase leading to a close final orbit.
The same scenario is also responsible for $\sim 5\,$\% of the close-orbit NSNS systems. This formation channel 
(direct-supernova) of close-orbit systems was first discussed by Kalogera~(1998) in the context of LMXBs.
Another comment to Fig.~\ref{merge_hist} is that the population of close-orbit NSBH systems is clearly seen to
be subdivided into two distinct populations.
The systems with $\tau _{\rm merge} < 9$~Gyr evolved through helium star RLO (case~BB) onto the NS whereas the systems
with  $9 < \tau _{\rm merge}/{\rm Gyr} <12$ evolved without helium star mass-transfer.
A close-up of the distribution of merging timescales $< 1$~Gyr is presented in Fig.~\ref{merge_hist1}.
The various populations have rather different distributions of $\tau _{\rm merge}$ -- ranging from
extremely short (NSBH) to somewhat flat (BHNS).
\begin{table*}
  \caption{Adopted standard parameters for our binary population synthesis code. A typical simulation is based on 20~million ZAMS binaries.}
  \label{stdpar}
  \begin{center}
  \begin{tabular}{llll}
  \hline
                Parameter & Symbol & Value & Note\\
  \hline
Mass of primary star             & $M_{\rm p}$               & $10-100\,M_{\odot}$     & Salpeter IMF (-2.7) \\
Mass of secondary star           & $M_{\rm s}$               & $4-M_p$                 & the distribution is drawn to match the q-distribution\\
Initial ZAMS separation          & $a_0$                     & $5-10\,000\,R_{\odot}$  & a flat distribution in $\log a$\\
Stellar wind mass loss           & $\alpha$                  & 0.20                    & \\
Timescale for RLO                & $\Delta t_{\rm rlo}$      & $\sim 3\,\tau_{\rm thermal}$ & generally a good estimate\\
Critical mass ratio for CE       & $q_{\rm ce}$              & 2.5                     & for a hydrogen donor star (at onset of RLO {\em after} wind mass loss)\\
Timescale for CE                 & $\Delta t_{\rm ce}$       & 1000~yr                 & upper limit\\
Efficiency of CE                 & $\eta_{\rm ce}$           & 0.5                     & + liberated $E_{\rm acc}$, see Eq.~\ref{Eacc}\\
$\lambda$ for CE                 & $\lambda$                 & $0.006\sim 0.4$         & individually calculated from stellar structure 
                                                                                         calculations, see Sec.~\ref{CE}\\
Core boundary parameter          & $f_{\lambda}=\lambda^{*}/\lambda$  & 2              & since ${X}\simeq 0.10$ may often underestimate the core mass 
                                                                                         (Tauris \& Dewi 2001)\\
Critical He-star mass for CE     & $M_{\rm He}^{\rm crit}$   & $3.3\,M_{\odot}$        & from recent helium donor star calculations (Dewi \& Pols 2003)\\
$\lambda$ for He-star CE         & $\lambda_{\rm He}$        & 0.1                     & typical value from (Dewi \& Pols 2003)\\
Maximum accretion rate           & $\dot{M}_{\rm max}$     & $\dot{M}_{\rm Edd}$       & an Eddington accretion limit is applied to both NS and BH\\
kick-distribution for NS         & $w$                     & $0-2000$~km~s$^{-1}$      & 3 comp. Maxwellian [$\sigma$=30 (5\%), 175 (80\%) and
                                                                                         700 km~s$^{-1}$ (15\%), respectively]\\ 
kick-distribution for BH         & $w_{\rm bh}$            & $0-2000$~km~s$^{-1}$      & same as above, but scaled: 
                                                                         $w_{\rm bh}=w\,(M_{\rm ns}/M_{\rm bh})$ to yield same linear momentum\\
  \hline
  \end{tabular}
  \end{center}
\end{table*}

\subsection{Galactic formation rates, merger rates and the present population of massive double degenerate binaries}
In Table~\ref{mergerrate_std} we list our best estimate for the Galactic formation rate, the merger rate
and the present population of massive double degenerate binaries. The present population in the Galactic
disk is a rough estimate based on the difference in formation rate and merger rate
integrated over the age of the Milky Way (12~Gyr) and corrected for: I) the systems escaping the
gravitational potential of our Galaxy ($<10\,$\%) and II) 
the contribution from the current population of potential mergers
recently formed. 
   \begin{table}
      \caption[]{The formation rates, the merger rates and the present population of massive double degenerate binaries
                 in the Galactic disk -- see text.}
         \begin{center}
         \begin{tabular}{cccr}
            \hline
            \noalign{\smallskip}
            Systems & Formation rate & Merger rate & $N_{\rm Gal}$\\ 
            \hline
            NSNS    & $1.8\times 10^{-6}$ yr$^{-1}$ & $1.5\times 10^{-6}$ yr$^{-1}$ & 3$\,$700\\  
            NSBH    & $6.3\times 10^{-7}$ yr$^{-1}$ & $8.4\times 10^{-8}$ yr$^{-1}$ & 6$\,$400\\  
            BHNS    & $1.1\times 10^{-6}$ yr$^{-1}$ & $5.0\times 10^{-7}$ yr$^{-1}$ & 6$\,$900\\  
            BHBH    & $3.7\times 10^{-5}$ yr$^{-1}$ & $9.7\times 10^{-6}$ yr$^{-1}$ & 320$\,$000\\  
            \hline
         \end{tabular}
         \label{mergerrate_std}
         \end{center}
   \end{table}
The formation rate of NSNS systems is larger than that of both NSBH and BHNS systems. However, as we noticed from
Fig.~\ref{merge_hist}, discussed in the previous section, these latter systems form a large fraction of
very wide-orbit binaries which will never merge and therefore accumulate in numbers with the age of the Galaxy.
This is also seen from their relatively small merger rates.
It is thus expected that the Galaxy hosts a larger number of NSBH and BHNS systems compared to NSNS systems.
The BHBH systems, however, are seen to dominate the overall Galactic population of massive compact binaries.
We estimate a present population above $300\,000$ systems in the Milky Way.
Although such systems cannot be detected observationally yet (since they do not emit electromagnetic radiation)
we may hope to detect continuous gravitational waves from some of these systems in our backyard once LISA has been 
launched.

\begin{figure}
   \centering
   \includegraphics[height=7cm]{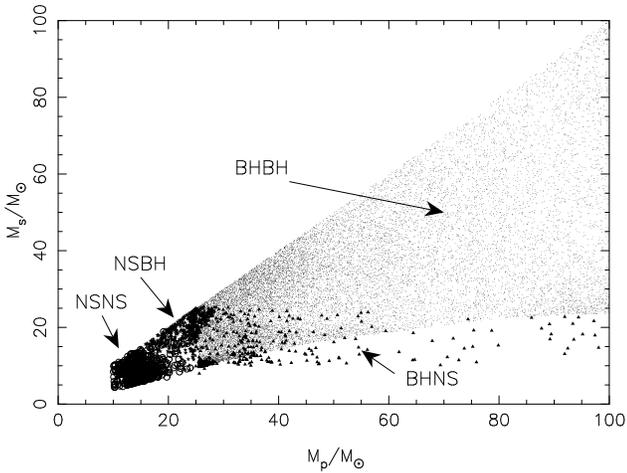}
      \caption{Initial primary and secondary stellar masses for 20$\,$000 ZAMS binaries evolving
               all the way to form massive double degenerate systems.}
       \label{initial}
   \end{figure}

In order to understand the large formation rate of BHBH binaries, and the origin of the different
massive compact binaries, it is usefull to look at Fig.~\ref{initial}. It is clearly seen that the initial
parameter space for the stellar component masses favour the formation of BHBH binaries. The lines separating
the different populations in the diagram can largely be understood from the threshold masses listed in
Table~\ref{Remnant}. The dividing lines are ``fuzzy'' since the initial orbital separation is also
a determining factor for the outcome of the binary stellar evolution (i.e. a helium star's mass depends on at what
evolutionary stage its progenitor star lost its hydrogen envelope). The lower curved line in the plot
marks the onset of common envelope evolution. The line is curved (rather than straight) since the
mass-ratio limit $q_{\rm ce}$ is evaluated {\em after} the wind mass loss is accounted for, prior
to the RLO, and more massive stars loose a relatively larger fraction of their mass in the form of a stellar wind
(see Fig.~\ref{HR}).
We notice that no BHBH systems are formed which evolved through a CE-phase during the
mass transfer from the primary star. The reason for this is the small values of the $\lambda$-parameter
obtained from real stellar structure models. This reflects that the envelopes of these massive stars
are tightly bound to their cores so that the binaries do not have enough orbital energy to expel the
envelopes in the spiral-in phase. Therefore the systems will merge (and perhaps form Thorne-\.Zytkow objects).

\subsection{Galactic double neutron star (NSNS) binaries}
\begin{figure*}
  \centering
 \includegraphics[height=10cm]{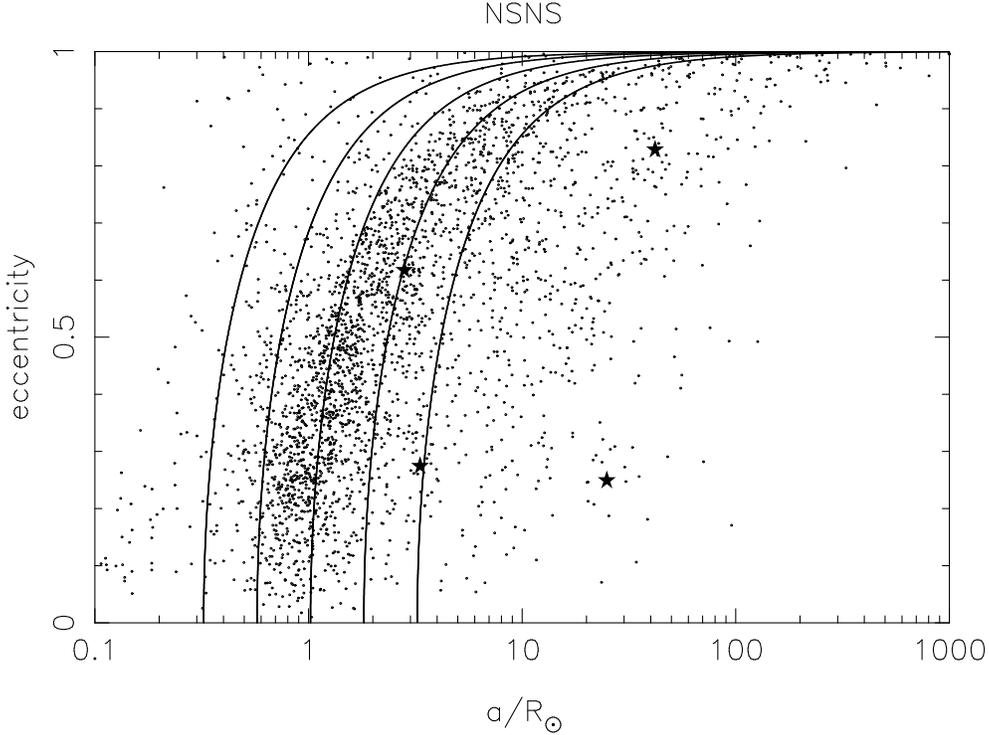}
  \caption{The simulated distribution of newborn double neutron star binaries in the (separation, eccentricity)-plane.
           Isochrones for the merging time of the double neutron star binaries are also shown.
           We calculated these assuming two $1.4\,M_{\odot}$ neutron stars in each binary.
           The curves correspond to values of: $3\times 10^5$~yr, 3~Myr, 30~Myr, 300~Myr and
           3~Gyr (from left to right), respectively. The four NSNS systems detected 
           in the Galactic disk are indicated with stars.} 
  \label{ae}
\end{figure*}
Simulating the formation of double neutron star binaries is a crucial test for the computer code and
the physics assumptions being used. In Fig.~\ref{ae} we have plotted the distribution of NSNS binaries
at birth in a (separation, eccentricity)-diagram. We have partly used the location of the four
known Galactic NSNS systems in the disk (see Table~\ref{NSNSobs}) to constrain our input parameters so we can
reproduce these observed systems from simulations. Especially PSR~J1518+4904 (Nice, Sayer \& Taylor~1996)
is a challenge for any binary population synthesis code given its low eccentricity (0.27) and relatively
wide orbit $P_{\rm orb}=8.6$~days ($a\simeq 25\,R_{\odot}$).
We find that our simulated population fits the observations well given that 
the majority of the simulated NSNS systems merge on a timescale much shorter than the lifetime of the observed
(recycled) pulsars. Furthermore, NSNS systems with $a\la 1\,R_{\odot}$ ($P_{\rm orb}\la 100$~min.) are very
difficult to detect as a result of the large acceleration of their pulsed signals.
It is possible that this plot may help to converge some of the many various input parameters used in population
synthesis codes today.
   \begin{table}
      \caption[]{The four observed double neutron star systems in the Galactic disk (for references see Nice, Sayer \& Taylor~1996 and Lyne et~al. 2000).}
         \label{NSNSobs}
         \begin{center}
         \begin{tabular}{lllcr}
            \hline
            PSR-name & sep.        & ecc. & $M=M_1+M_2$ & $\tau\equiv P/2\dot{P}$\\
                     & $R_{\odot}$ &      & $M_{\odot}$ & Myr\\
            \hline
            J~1518+4904 & 24.9 & 0.249 & 2.69 & 20$\,$000\\
            B~1534+12   & 3.33 & 0.274 & 2.68 & 200\\
            J~1811--1736& 41.9 & 0.828 & 2.6  & 900\\
            B~1913+16   & 2.81 & 0.617 & 2.83 & 100\\
            B~2127+11C$^{\mathrm{*}}$  & 2.86 & 0.681 & 2.71 & 100\\
            \hline
         \end{tabular}
\begin{list}{}{}
\item[$^{\mathrm{*}}$] This system resides inside a globular cluster and has a different origin. 
\end{list}
         \end{center}
   \end{table}

\begin{figure}
   \centering
   \includegraphics[height=7cm]{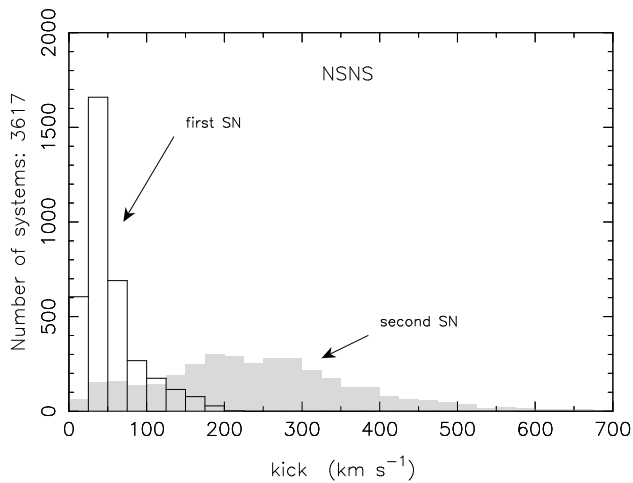}
      \caption{The distribution of kicks imparted to newborn neutron stars in {\em surviving} binaries.
               The difference in selected kicks between the first and second SN are expected from
               a binary evolutionary point of view -- see text.}
      \label{NSNSkicks}
   \end{figure}

It is instructive to look at the distribution of kicks imparted to neutron stars in surviving binaries which
evolved all the way to the final NSNS stage. This is shown in Fig.~\ref{NSNSkicks}.
The difference between the kicks imparted during the first and the second SN explosion is very significant:
the average values of the selected kicks are $\bar{w}_1=53$~km~s$^{-1}$ and $\bar{w}_2=263$~km~s$^{-1}$,
respectively.
In order for the systems to avoid merging in the CE and spiral-in phase of the evolved secondary stars, the
systems must be in a fairly wide orbit prior to the HMXB and subsequent CE and spiral-in phase.
This means that these systems can only survive very small kicks in the first SN explosion in order to avoid disruption.
On the other hand, after the CE and spiral-in phase the orbits of the surviving systems will always be tight
and therefore the systems will be able to withstand even very large kicks in the second SN explosion.
Actually, the distribution of selected kicks during the second SN resembles the input trial distribution
of kicks plotted in Fig.~\ref{kick}. The corresponding systemic recoil velocities after the SNe were
$\bar{v}_1=8.6$~km~s$^{-1}$ and $\bar{v}_2=220$~km~s$^{-1}$, respectively. Thus, when considering the runaway
properties of NSNS systems it is sufficient to consider the effects from the second SN.

\begin{figure*}
  \centering
 \includegraphics[height=10cm]{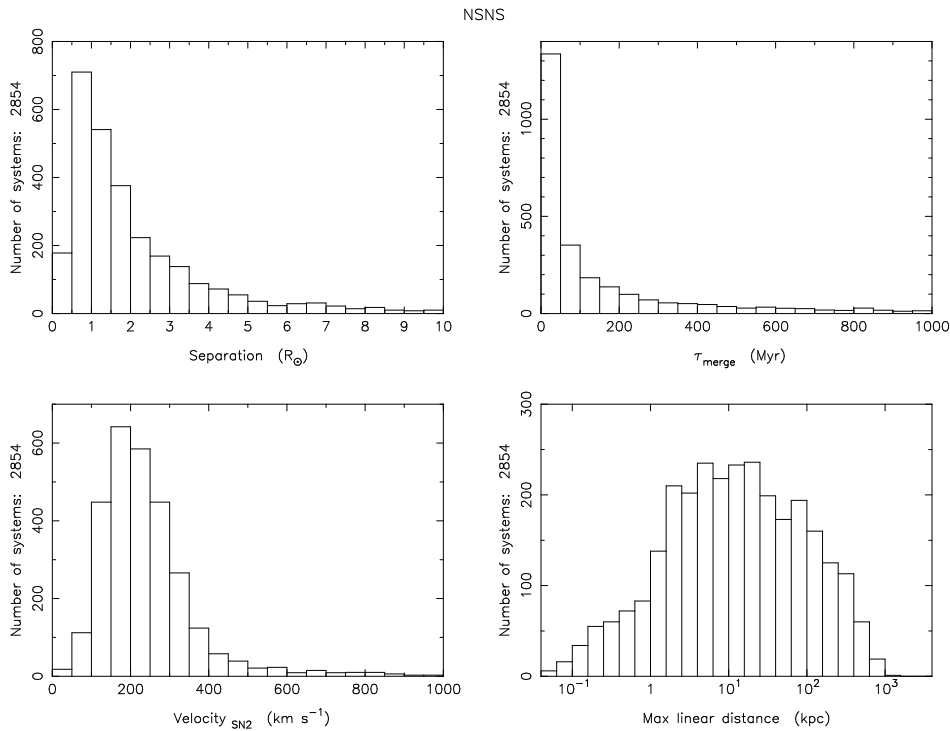}
  \caption{Distribution of newborn double neutron star binaries with respect to orbital 
           separation, merging timescale, systemic (recoil) velocity from the second SN explosion 
           and the maximum distance from the origin of birth in a host galaxy -- see text for details.}
  \label{NSNShisto}
\end{figure*}
To get an idea of the distance travelled by NSNS binaries from their birth place prior to their coalescence,
one simply has to multiply their space velocity with the merging timescale. The space velocity is caused
by the recoil of shell ejection combined with the asymmetric kick given to the neutron star at birth.
In Fig.~\ref{NSNShisto} we have plotted histograms of final orbital separation, merger time, space velocity 
and the estimated distance travelled before the merging event. In order for other readers to compare with
our results we have in this plot simplified both the space velocity and the distance travelled.
For the space velocity shown here it is assumed that the binary was at rest prior to the second SN (which
is a good approximation, see above).
The other approximation applied here is that the binary motion was without influence from the
Galactic potential (this approximation gets exceedingly worse with increasing merger times). Hence, the
estimated distance is an upper limit.
We see that typical systems travel $\sim$10~kpc prior to their merging and that the distribution of
distances travelled is rather broad. 
It should be mentioned that in our simulations presented throughout this paper 
the space velocities
resulting from both SNe are added as vectors and the binary is moving in a Galactic potential
as described in Section~3.

\subsection{Offset distribution of mergers from their host galaxies}
\begin{figure*}
  \centering
 \includegraphics[height=22cm]{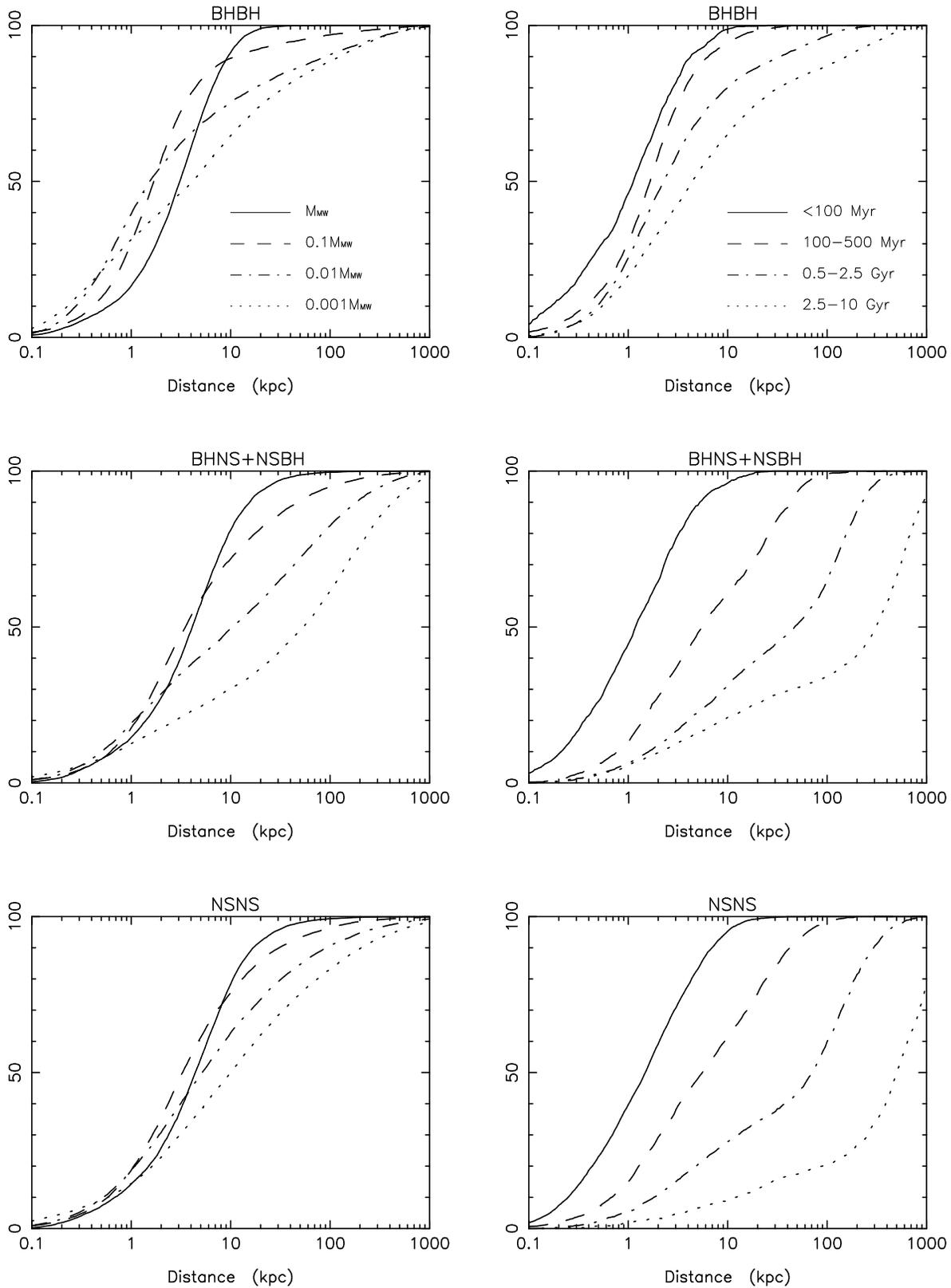}
  \caption{The panels in the left column shows the cumulative distribution of systems merging within the projected
           radial galactic distance given on the x-axis. Only binaries merging within 10~Gyr are included.
           We launched our binaries into four galaxies with different
           masses (0.001--1 times the mass of the Milky Way). 
           The panels on the right-hand side are for 
           a $0.1\,M_{\rm MW}$ galaxy. The curves represent four different merging timescale intervals -- see text.}
  \label{gal}
\end{figure*}
Recent work by Bloom, Kulkarni \& Djorgovski (2002) shows that the median projected angular
offset of 20 long-duration GRBs is only $\sim$1.3~kpc (0.$\!$"17 for an assumed cosmology of
$\Omega _\Lambda = 0.7, \Omega _M = 0.3$ and $H_0=65$~km~s$^{-1}$~Mpc$^{-1}$). In Fig.~\ref{gal} we have plotted
our cumulative distribution of calculated projected radial distances in various host galaxies.
For a random orientation of the radial distance with respect to the line-of-sight one finds the average projected 
radial distance (in the plane of the sky) by multiplying with a factor~$\pi/4$.
The panels in the left column show the result of launching our binaries into four galaxies with different
masses (0.001--1 times the mass of the Milky Way). We see that roughly half of
all massive compact binaries  merge within a projected radial distance of 4 kpc for a $0.1\,M_{\rm MW}$ galaxy. 
Hence, the majority of these systems merge {\em inside} their host galaxies. However, the median distance
is larger than the median value observed for the
long-duration bursts.  At a projected radial offset of 1.3~kpc we find that less than 25\% of all BHNS, NSBH and NSNS
systems merge (independent of the mass of the host galaxy). 
In general, it therefore seems clear that merging compact objects most likely cannot be the progenitors of 
long-duration GRBs (as also concluded by Bloom, Kulkarni \& Djorgovski 2002).
However, from a pure kinematical point of view one cannot rule out that some of the long GRBs with afterglows could
be associated with merging compact objects. It is also an interesting question to what extent such a merging event
outside a galaxy would produce an afterglow at all.
It should be noticed that the merging BHBH systems are only shown for comparison -- they
cannot be progenitors of GRBs. Note, that the central panel shows the mixed population of both NSBH and BHNS systems.
The panels on in the right-hand side of the figure simply show how the cumulative fraction of systems merging increases
as a function of projected radial galactic
distance for a $0.1\,M_{\rm MW}$ galaxy. The curves represent four different merging timescale intervals:
0--100~Myr, 100--500~Myr, 0.5--2.5~Gyr and 2.5--10~Gyr, respectively. 
We find that $\sim$10$\,$\% of all NSNS systems formed escape our Galaxy.

\begin{figure}
  \centering
 \includegraphics[height=7cm]{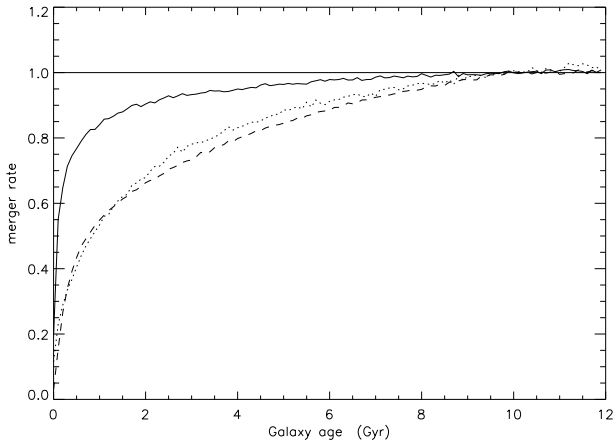}
  \caption{Merger rates (relative to present merger rates) as a function
           of the age of our Galaxy for NSNS systems (solid line),
           BHNS+NSBH systems (dashed line) and BHBH systems (dotted line) for a 
           constant Galactic star formation rate. 
           The curves show the emergence of a ``steady state''.}
  \label{gal2}
\end{figure}

Fig.~\ref{gal2} shows the merger rates of NSNS systems (solid line),
BHNS+NSBH systems (dashed line) and BHBH systems (dotted line) for a 
constant Galactic star formation rate. Each of the merger rates are scaled relative
to its present merger rate.
In calculating this rate, only binaries merging within the age of the Milky Way are included.
It is seen that the Milky Way today is in a ``steady state'' (i.e. the formation rate of systems in close orbits merging
within 10~Gyr is equal to the number of systems merging per unit time).  

\subsection{Prospects for LIGO/VIRGO detection rates}
LIGO~I and LIGO~II are expected to detect
NSNS inspiral events out to a distance of $\sim 20$~Mpc and $\sim 300$~Mpc, respectively, according
to recent estimates (Thorne~2001).
This corresponds to wave amplitudes of roughly $10^{-20} > h > 10^{-22}$. As a result of the much larger
chirp~mass ($M_{\rm chirp}\equiv \mu ^{3/5}M^{2/5}$) of the BHBH mergers, such binaries will be detected out to a distance luminosity,
$d_L \propto M_{\rm chirp}^{5/6}$ (Finn 1998) which is about 4 times larger in average according to our estimates. 
Hence, the ratio of detected event rates for BHBH mergers relative to NSNS mergers is
$4^3 \sim$ 64 times larger than the corresponding ratio of such mergers in our Galaxy. Thus BHBH mergers
will be completely dominant for LIGO detections (as noted by Sipior \& Sigurdsson 2002). 
The exact
value of the chirp mass ratio depends mainly on the black hole masses estimated at birth as well as the possibility
of hypercritical accretion onto neutron stars.
BHNS and NSBH mergers have a chirp mass which is roughy 2~times larger than that of the NSNS systems and thus the corresponding detected
ratio of such mergers, relative to NSNS mergers, is $\sim$5 times larger than the Galactic ratio.

In order to extrapolate the Galactic coalescence rate out to the volume of the universe accessible
to LIGO, one can either use a method based on star formation rates, galaxy number density or a scaling based on the
B-band luminosities of galaxies. Using the latter method Kalogera~et~al.~(2001) found a scaling
factor of $(1.0-1.5)\times10^{-2}$~Mpc$^{-3}$, or equivalently,
$\sim 400$ for LIGO~I (out to 20~Mpc for NSNS mergers). Since LIGO~II is expected to look out to a distance
of 300~Mpc (NSNS mergers), the volume covered by LIGO~II is larger by a factor of (300/20)$^3$ and thus the
scaling factor in this case, relative to the coalescence rates in the Milky Way, is about $1.3\times 10^6$.
Therefore, the expected rate of detections from galactic field NSNS inspiral events is roughly 2~yr$^{-1}$
for LIGO~II (see Table~\ref{LIGO}) and an impressive 840~yr$^{-1}$ for the BHBH mergers! 
Even LIGO~I may, with a bit of luck, detect an event from a BHBH collision.
   \begin{table}
      \caption[]{The expected LIGO/VIRGO detection rates of compact mergers.}
         \begin{center}
         \begin{tabular}{cccr}
            \hline
            \noalign{\smallskip}
            Systems & Galactic merger rate & LIGO~I & LIGO~II\\ 
            \hline
            NSNS    & $1.5\times 10^{-6}$ yr$^{-1}$ & $6.0\times 10^{-4}$ yr$^{-1}$ & 2.0 yr$^{-1}$\\ 
            NSBH    & $8.4\times 10^{-8}$ yr$^{-1}$ & $1.7\times 10^{-4}$ yr$^{-1}$ & 0.6 yr$^{-1}$\\
            BHNS    & $5.0\times 10^{-7}$ yr$^{-1}$ & $1.0\times 10^{-3}$ yr$^{-1}$ & 3.4 yr$^{-1}$\\ 
            BHBH    & $9.7\times 10^{-6}$ yr$^{-1}$ & $2.5\times 10^{-1}$ yr$^{-1}$ & 840 yr$^{-1}$\\
            \hline
         \end{tabular}
         \label{LIGO}
         \end{center}
   \end{table}
Beware, that our estimated merger rates should be considered as lower limits given that compact mergers
in globular clusters probably also contribute significantly to the total merger rates (Portegies Zwart \& McMillan 2000).
The cosmological implications of gravitational
wave observations of binary inspiral are also interesting to note (e.g. Schutz 1986; Finn 1997).

\section{Discussion}
\subsection{The dependence on input parameters}
Monte Carlo simulations is a powerful tool for investigating complicated interactions which depend on many parameters.
Often the difficult task is not to produce a reasonable code but to analyse the results, pinpoint the important
physical parameters behind the major trends in a simulated population and try to understand these results
in terms of relatively simple physics.
   \begin{table}
      \caption[]{Galactic relative formation ratios, formation rates and merger rates of massive compact binaries.
                 Model~{A} is our best estimate. See text.}
         \begin{center}
         \begin{tabular}{lccc}
            \hline
             Model/ & Rel. formation & Formation rate & Merger rate\\
             Systems        & \%             &  Myr$^{-1}$    & Myr$^{-1}$\\
            \hline
            A) standard & & &\\
            \cline{1-1}
            NSNS   & 4.4  & 1.8  & 1.5\\ 
            NSBH   & 1.5  & 0.63 & 0.08\\ 
            BHNS   & 2.7  & 1.1  & 0.50\\
            BHBH   & 91.3 & 37   & 9.7\\
            \hline
            B) $\lambda =0.5$ constant & & &\\
            \cline{1-1}
            NSNS   & 7.6  & 20  & 17\\ 
            NSBH   & 2.3  & 6.1 & 3.7\\ 
            BHNS   & 1.9  & 4.9 & 1.3\\
            BHBH   & 88.2 & 230 & 76\\
            \hline
            C) kick: $0.5\,\sigma _w$ & & &\\
            \cline{1-1}
            NSNS   & 6.5  & 7.2  & 3.3\\ 
            NSBH   & 3.6  & 4.0  & 0.34\\ 
            BHNS   & 3.4  & 3.8  & 1.3\\
            BHBH   & 86.5 & 96   & 18\\
            \hline
            D) core: $f_\lambda =1$ & & &\\
            \cline{1-1}
            NSNS   & 3.6  & 1.1  & 0.95\\ 
            NSBH   & 1.6  & 0.48 & 0.04\\ 
            BHNS   & 2.9  & 0.61 & 0.16\\
            BHBH   & 92.7 & 27   & 1.3\\
            \hline
            E) IMF: $f\propto M_p^{-2.0}$ & & &\\
            \cline{1-1}
            NSNS   & 4.4  & 1.9  & 1.4\\ 
            NSBH   & 1.6  & 0.66 & 0.08\\ 
            BHNS   & 2.9  & 1.2  & 0.54\\
            BHBH   & 91.0 & 37   & 9.8\\
            \hline
            F) wind: $\alpha=0$ & & &\\
            \cline{1-1}
            NSNS   & 2.8  & 1.1  & 0.83\\ 
            NSBH   & 3.5  & 1.4  & 0.22\\ 
            BHNS   & 2.0  & 0.77 & 0.22\\
            BHBH   & 91.8 & 36   & 7.2\\
            \hline
            G) $q_{\rm ce}=2.0$ & & &\\
            \cline{1-1}
            NSNS   & 3.6  & 1.3  & 1.0\\ 
            NSBH   & 1.5  & 0.54 & 0.08\\ 
            BHNS   & 3.7  & 1.3  & 0.44\\
            BHBH   & 91.2 & 32   & 6.9\\
            \hline
            H) $q_{\rm ce}=3.0$ & & &\\
            \cline{1-1}
            NSNS   & 5.2  & 2.4  & 1.9\\ 
            NSBH   & 1.5  & 0.71 & 0.11\\ 
            BHNS   & 3.7  & 1.7  & 1.1\\
            BHBH   & 89.5 & 41   & 12\\
            \hline
            I) $q_{\rm ce, He}=2.5$ & & &\\
            \cline{1-1}
            NSNS   & 2.3  & 0.93  & 0.54\\ 
            NSBH   & 1.5  & 0.62 & 0.02\\ 
            BHNS   & 2.9  & 1.2  & 0.57\\
            BHBH   & 93.3 & 38   & 9.7\\
            \hline
            J) $\Delta E_{\rm acc}=0$ & & &\\
            \cline{1-1}
            NSNS   & 4.2  & 1.4 & 1.3\\ 
            NSBH   & 2.2  & 0.71 & 0.06\\ 
            BHNS   & 2.2  & 0.71 & 0.21\\
            BHBH   & 91.4 & 30   & 3.4\\
            \hline
            average (D$\rightarrow$J)& & &\\
            \cline{1-1}
            NSNS   & 3.7  & 1.5  & 1.1\\ 
            NSBH   & 1.9  & 0.74 & 0.09\\ 
            BHNS   & 2.9  & 1.0  & 0.47\\
            BHBH   & 91.6 & 34   & 7.2\\
            \hline
         \end{tabular}
         \label{input}
         \end{center}
   \end{table}

In Table~\ref{input} we summarize
the effects of various input parameters on the formation and evolution of massive double degenerate systems.
Model~{A} is our standard model. In all other models we modified one parameter at a time, keeping all the other
parameters as in our standard model. 
Model~{B} having a constant $\lambda =0.5$ is a very bad approximation (see Fig.~\ref{lambdafig}), but nevertheless
it is apparently still used in many computer codes. For example it is clear that the formation channel recently
presented in Fig.~1 of Belczynski, Bulik \& Kalogera (2002) will not work if one applies a reasonable realistic
value for the binding energy of the envelope. At stage~{V} in their figure, one can use a stellar structure model and
estimate a value of $\lambda =0.23$ 
for the $R=82\,R_{\odot}$ donor star of mass $14.1\,M_{\odot}$. This results in a post-CE separation which is roughly
3~times {\em smaller} than the radius of the descendant helium star. Hence, this system is doomed to coalesce (if one excludes
the internal thermodynamic energy when estimating the value of $\lambda$ the problem even exacerbates).
Model~{C} seems to give too small values of the kick velocity
compared to observations of radio pulsars (Cordes \& Chernoff 1998). At the bottom of the table the average result of models~{D}--{J} is given.
It is interesting to notice that these values are not very different from our standard model results.

\subsection{The Galactic supernova rate}
We have based the calibration of our models on the star formation rate for the Milky Way (see Eq.~\ref{BSFR}). 
In order to check the validity of this calibration we have compared our simulated Galactic supernova rate
of type~Ib/c SNe with the empirical estimates of Cappellaro~et~al. (1999). If we assume that all type~Ib/c SNe originate
from the collapse of naked stars which have lost their hydrogen/helium envelopes as a result 
of mass transfer in a binary system, then we find a rate of: $f\,0.28$~(100~yr)$^{-1}$. We can compare this result
with a rate of 0.14~SNu (1~SNu = 1~SN (100~yr)$^{-1}$ $(10^{10}\,L_{\odot}^B)^{-1}$)
given in Table~4 in Cappellaro~et~al. (1999). The B-band luminosity of the Milky~Way is $\sim$ a few times
$10^{10}\,L_{\odot}$ and therefore our calibration seems to be about right.

\subsection{Comparison with other studies}
The expected LIGO/VIRGO detection rates can be determined either from binary population
synthesis calculations, or from observations of Galactic NSNS systems (binary pulsars).
Both methods involve a large number of uncertainties -- see the excellent review by Kalogera~et~al.~(2001) and references therein.
Current estimates for the Galactic merger rate of NSNS systems converge between $10^{-6} - 10^{-4}$~yr$^{-1}$.
Our results belong to the lowest end of this interval. However, notice from Table~\ref{input}
that if ({\em when}) other population synthesis codes start using real $\lambda $-values (rather than a constant) their merger rates
will go down by a factor of $\sim$10. 

In general there is no consensus on the formation ratios of the various compact binaries.
This is due to the $\lambda$-values discussed in this paper,
the helium star evolution and
the unsettled question of hypercritical accretion. Studies including
hypercritical accretion adopt different approaches and do not agree on the
extend and outcome of this effect. Studies including this process
turn many of the (first-born) neutron stars into black holes, resulting in
a much larger ratio of BHNS/NSNS.
In our study we find a relatively high
number of BHBH systems being formed. We wonder if other codes reject
systems which do not evolve through a (semi-detached) contact phase (i.e.
if the primary star evolves to giant dimensions without filling its
Roche-lobe). As we have seen in this study, the direct-supernova
mechanism (Kalogera 1998) forms a significant fraction ($\sim$1/3) of
all close-orbit BHNS systems and also contributes ($\sim$5\%) somewhat
to the formation of NSNS mergers.

The distribution of separations (or lifetimes) of the merging binaries
is roughly in agreement with the results of other recent studies. The double
neutron stars in the study of Belczynski, Kalogera \& Bulik (2002) have
slightly tighter orbits. This is mainly due to their higher limit of $M_{\rm He}^{\rm crit}$
resulting in more systems going through a second common envelope phase. Their
BHNS systems are marginally wider than ours, since their progenitors all are
more massive than this limit. The distribution of NSNS separations found by
Bloom, Sigurdsson \& Pols (1999) is close to ours, even though they have
not included mass transfer from a helium star. This is somewhat a surprise since
it is required that the helium stars lose their envelope in order to form very
tight orbits without merging.

We find our distribution of space velocities of the merging binaries
to closely resemble both that of Belczynski, Kalogera \& Bulik (2002), and
that of Bloom, Sigurdsson \& Pols (1999).
Our distribution of merger sites mostly rely on the lifetimes and space 
velocities, and is therefore also in agreement with their studies.

Our parameter variation studies show that the distributions of lifetimes
and space velocities of the merging binaries are quite stable (unlike the formation rates).
This might seem somewhat surprising. If, as an example,
the inspiral of the neutron star during the common envelope is more
efficient, then it would seem that the distribution of NSNS systems should
become tighter. This does not happen due to the fact that only a limited
range of post-inspiral distances lead to survival of NSNS systems. If the orbit becomes
too tight the neutron star merges with the core of the giant. If, on the other hand, the
orbit is too wide then the supernova explosion is likely to disrupt the
binary.

\subsection{Progenitors of short-duration GRBs}
As discussed in the introduction, it seems likely that one needs a disk of nuclear matter around a black hole
in order to produce a GRB. Such a disk can be formed by the collapse of the core of a massive star or by a merger
of either a double neutron star or a mixed neutron star/black hole system. The merger of a black hole and a white dwarf
(Fryer et al. 1999),
or the merger of a neutron star or black hole with a {CO}- or helium core (Fryer \& Woosley 1998), as a result of spiral-in in a CE,
may not be a good candidate for producing a short-duration GRB. One concern is that a white dwarf, or a {CO}/He-core,
is a relatively low-density object compared to nuclear matter and it has a relatively large dimension ($>5000$~km).
The resultant disk would probably then be accreted much slower than needed for making a short-duration GRB.
Merging double black holes are not candidates either due to the complete lack of any accretion disk. Therefore, we only
considered the merging of double neutron stars (NSNS) or neutron star/black holes (BHNS and NSBH) as progenitors
of the short-duration GRBs. 

\subsection{Do all GRBs origin from binaries ?}
It is interesting to notice that most fireball models eject only a little matter ($\sim 10^{-5}\,M_{\odot}$)
and require a rotating progenitor. To minimize the amount of ejected matter, collapsars require that the
hydrogen envelope of the massive star has been removed before the collapse (MacFadyen \& Woosley 1999). 
So the progenitors are in fact
rotating helium stars (Wolf-Rayet stars) collapsing into black holes. Strong winds could remove the hydrogen
envelope but may also remove too much spin angular momentum. An effective way of removing the envelope is via
RLO in a binary system. Hence, it is possible that {\em all} GRBs origin from massive interacting binaries.
In a recent paper (Lee, Brown \& Wijers~ 2002) it has been suggested that indeed collapsars (hypernovae)
are formed in tight binaries and that soft {X}-ray transients (SXTs) with black hole accretors are the
{\em descendants} of GRBs. 

\subsection{Hypercritical accretion}
It has been argued that a neutron star engulfed in a common envelope might experience
hypercritical accretion and thereby collapse into a black~hole
(e.g. Chevalier~1993; Brown~1995). However, this idea seems difficult to conciliate with
the observations of a number of very tight-orbit binary pulsars (Tauris, van~den~Heuvel \& Savonije 2000). For example, the system
PSR J1756--5322 which has a {CO} white dwarf companion ($\sim \!0.7\,M_{\odot}$) with an orbital period
of only 0.45~days. From an evolutionary point of view there is currently no other way to produce such a
system besides from a CE and spiral-in phase.
From a theoretical point of view it has been argued that the hypercritical accretion can be
inhibited by rotation (Chevalier 1996) and strong outflows from the accretion disk (Armitage \& Livio 2000).
A number of other population synthesis studies have included hypercritical accretion as always leading
to collapse of the neutron star (e.g. Fryer, Woosley \& Hartmann 1999), or as an intermediate approach in which
collapse only happens if the neutron star accretes a certain amount of matter (e.g. Belczynski, Bulik \& Kalogera 2002).
In the latter case a new parameter, the critical mass for a neutron star to collapse into a black hole
is introduced (depending on the equation of state of neutron stars).
Note however, that the masses of neutron stars determined in all of the five detected double
neutron star systems are quite similar and close to a value of $1.4M_{\odot}$.
Hence, in all these cases it is clear that the first-born neutron star did not accrete any
significant amount of material from the common envelope. For the reasons mentioned above,
we currently refrain from the possibiliby of hypercritical accretion onto
neutron stars which successfully eject the envelope of their companion star.

\section{Conclusions}
We have performed binary population synthesis 
in order to estimate the properties
of massive compact binaries. In particular we have
focused on the kinematics of close-orbit binaries which merge within
a Hubble-time as a result gravitational wave radiation.
Our simulated populations of massive double degenerate systems are relatively
stable against variation of the parameters of the code.

   On a number of points we find that our description is 
   more advanced than the work previously published
   in the literature: 
\begin{itemize}
    \item We have used realistic values of the $\lambda$-parameter of the
    \item[] common envelope and spiral-in evolution. We estimated this 
    \item[] parameter for each individual binary donor star.
   \item We have used realistic models for the evolution of helium stars
   \item[] and included the effects of mass transfer from helium stars.
  \item We have included the effects of energy feedback from an 
   \item[] accreting compact object in the common envelope evolution.
   \item We have applied reduced kicks to black holes in accordance 
   \item[] with observations of black hole binaries.
\end{itemize}

        We find that the formation rate and merger rate of massive double degenerate binaries
        is dominated by double black holes systems. The merger rates and the number of 
        such systems present in our Galaxy today are estimated to be (based on model~A):
\begin{itemize}
  \item The Galactic merger rate of BHBH systems is $\sim 9.7\times 10^{-6}$~yr$^{-1}$
  \item[] and the present disk population is $\sim 320\,000$ systems
  \item The Galactic merger rate of BHNS systems is $\sim 5.0\times 10^{-7}$~yr$^{-1}$
  \item[] and the present disk population is $\sim 6\,900$ systems
  \item The Galactic merger rate of NSBH systems is $\sim 8.4\times 10^{-8}$~yr$^{-1}$
  \item[] and the present disk population is $\sim 6\,400$ systems
  \item The Galactic merger rate of NSNS systems is $\sim 1.5\times 10^{-6}$~yr$^{-1}$
  \item[] and the present disk population is $\sim 3\,700$ systems
\end{itemize}
        The reason for the relatively low value of e.g. the NSNS merger rate is a result of our use of realistic
         $\lambda $-values.
        The mixed neutron star/black hole binaries have somewhat lower formation rates
        than the NSNS binaries, but a slightly larger population present in the Milky Way today,
        as a result of their longer merging timescales. Our studies also reveal an accumulating numerous 
        population of very wide-orbit BHBH systems which never merge.

        We estimate the detection rate for LIGO~II and find that this rate is totally
        dominated ($>99\,$\%) by BHBH mergers:
\begin{itemize}
    \item $\sim $840~yr$^{-1}$ for the BHBH mergers
    \item $\sim $3.4~yr$^{-1}$ for the BHNS mergers
    \item $\sim $0.6~yr$^{-1}$ for the NSBH mergers
    \item $\sim $2.0~yr$^{-1}$ for the NSNS mergers
\end{itemize}
        With a bit of luck LIGO~I may detect a merging event of a BHBH system (0.25~yr$^{-1}$).
        The numbers quoted above are from model~A.
        For other reasonable choices of parameter values these numbers typically differ within a factor
        of $\sim\! 2$ (see Table~\ref{input}).

        We have also explored the movement of the binaries in the gravitational
        potential of galaxies. This allows us to estimate the distribution of
        merging offsets from the centers of the galaxies. We find that a large
        fraction of the compact binaries are retained within the massive
        galaxies, but for lower galaxy masses these binaries are able
        to escape and merge outside the galaxies. This result is not consistent
        with the observed offset distribution of long-duration GRB afterglows. We therefore conclude
        that merging compact objects cannot account for the ensemble of GRB offset distributions observed today 
        -- e.g. simulated NSNS binaries have a
        median projected radial (offset) distance of $\sim 4$~kpc with respect to their host galaxy, while the observed
        value from 20 long-duration GRBs is only $\sim 1.3$~kpc.

\section*{Acknowledgments}
  We thank Onno Pols for providing helium star models. We also thank
  Jasinta Dewi and Ed van~den~Heuvel for discussions.
  T.M.T gratefully acknowledges support from the Danish Natural Science Research Council
  under grant no. 56916.

\end{document}